\documentclass[aps, prc, 12pt, nofootinbib, showpacs, superscriptaddress, tightenlines, groupedaddress]{revtex4-2}
\usepackage{amsmath,amssymb,amsbsy,bm}
\usepackage{graphicx}
\usepackage{comment}
\usepackage{float}
\usepackage[colorlinks=true,linkcolor=blue,citecolor=blue,urlcolor=blue]{hyperref}
\usepackage[margin=0.75in]{geometry}
\usepackage{silence}
\begin{document}

\title{Using Baryonic Charge Balance Functions to Resolve Questions about the Baryo-Chemistry of the QGP}
\author{Scott Pratt}
\affiliation{Department of Physics and Astronomy and Facility for Rare Isotope Beams\\
Michigan State University, East Lansing, MI 48824~~USA}
\author{Dmytro Oliinychenko}
\affiliation{Institute for Nuclear Theory\\
University of Washington, Seattle, WA 98195~~USA}
\author{Christopher Plumberg}
\affiliation{Illinois Center for Advanced Studies of the Universe, Department of Physics, University of Illinois at Urbana-Champaign, Urbana, IL 61801, USA}
\date{\today}

\pacs{}

\begin{abstract}
Baryon annihilations during the hadronic stage of heavy-ion collisions affects final-state baryon and antibaryon yields and final-state correlations of baryons and antibaryons. Understanding annihilation is important for addressing questions about the chemistry at the beginning of the hadronic stage, and for interpreting charge-balance correlations involving baryons. Here, charge balance functions, using protons and antiprotons binned by relative momentum, rapidity and azimuthal angle, are shown to clarify the amount of annihilation in the hadronic stage. This enables a more accurate extraction of the baryo-chemistry at the beginning of the hadronic stage. Understanding annihilation is also crucial if charge balance correlations are to be used to infer the chemistry of the earliest stages of a heavy-ion collision. Calculations are presented based on microscopic  simulations of the hadronic stage coupled to a hydrodynamic description of the earlier stage, along with a detailed modeling of correlations of protons and antiprotons, known as charge-balance functions.
\end{abstract}

\maketitle

\section{Introduction}\label{sec:intro}

As super-hadronic matter in a heavy-ion collision cools and hadronizes, it is common to assume that chemical compositions freeze-out close to the hadronization temperature, $T_H\approx 155$ MeV, with the yields at that point corresponding to chemical equilibrium. Of course, this is only approximate. Even if chemical equilibrium was valid at $T_H$, particles undergo additional interaction in the hadronic stage. One class of such interactions is baryon annihilation. Protons and antiprotons annihilate with cross sections that become large at lower invariant mass. A fit to data based on annihilation in antiproton beam physics \cite{Bruckner:1989ew,OBELIX:1996pze} gives the parameterization \cite{Wang:1998sh},
\begin{eqnarray}
\label{eq:sigmaA}
\sigma_{\rm A}&=& \frac{67}{P_{\rm lab}^{0.7}}~{\rm mb},
\end{eqnarray}
where $P_{\rm lab}$ is the momentum of a baryon in the rest frame of the other baryon in GeV/$c$. The annihilation cross section can become quite large. For $P_{\rm lab}<200$ MeV/$c$, cross sections exceed 200 mb.

Estimates of the amount of baryon annihilation vary from $\approx 10\%$ to $\approx 30\%$, with some of the variation depending on what type of model is being applied, and especially on whether regeneration is included \cite{Rapp:2000gy,Steinheimer:2012rd,Pan:2014caa,Seifert:2017oyb}. A typical two-proton annihilation might produce five pions. At equilibrium, or immediately after hadronization, the inverse process, $5\pi\rightarrow p,\bar{p}$, occurs with exactly the same rate as the annihilation \cite{Rapp:2000gy}. As the system cools and chemical equilibrium is lost, the regeneration rate is expected to fall well below the annihilation rate, with regeneration being rather important at the very final stages of the collision \cite{Pan:2014caa}. Thus, both annihilation and regeneration need to be considered. The role of annihilation has recently become more important given that the ALICE Collaboration at the LHC has reported that the $p/\pi$ ratio falls by $\approx 15-20\%$ from semi-central to the most central collisions \cite{ALICE:2013wgn}. Given that larger systems last longer and provide more opportunity for annihilation, one might wonder whether this reduction is partly due to additional annihilation in the hadronic phase.

Baryon annihilation is also of critical interest in the studies of charge-balance functions (BFs), which have been measured at both RHIC (Relativistic Heavy Ion Collider) and the LHC \cite{Westfall:2004jh,STAR:2015ryu,Wang:2011za,Wang:2011bea,Li:2011zzx,STAR:2010plm,Westfall:2004cq,STAR:2003kbb,STAR:2003kbb,Wang:2012jua,Abelev:2010ab,Adams:2003kg,Aggarwal:2010ya,Adamczyk:2015yga,STAR:2011ab,ALICE:2021hjb,Pan:2018dsq,Alam:2017iom,Weber:2013fla,ALICE:2013vrb,Weber:2012ut,Abelev:2013csa,Alt:2004gx,Alt:2007hk}. Baryonic charge must be locally accompanied by opposite charge. If a chemically-equilibrated quark-gluon plasma is created early in a heavy-ion collison, baryonic charge, quantified by the baryonic susceptibility, is created early (within the first fm/$c$), which leads to large separations in relative rapidity of balancing baryonic charges, e.g. protons and antiprotons. BFs, defined below, provide a measure of the separation of balancing charge \cite{Pratt:2017oyf}. For example, if a proton is observed in the detector, the BF represents the distribution of additional antiprotons vs. protons relative to the observed proton. If the $p\bar{p}$ BF is broad in relative rapidity, it would signal that chemical equilibrium was established early in the collision \cite{Pratt:2021xvg}. The proton-antiproton BF, when binned by relative azimuthal angle, also plays a pivotal role in extracting the light-quark diffusivity from experiment \cite{Pratt:2019pnd}. However, the shape of the BF binned by relative rapidity or azimuthal angle should also be affected by annihilation in the hadronic phase. Thus, for studying the diffusivity and chemical evolution of matter in a heavy-ion collisions, it is essential to understand how annihilation distorts the proton-antiproton BF.

In this paper, we illustrate how experiment can clarify the amount of baryon annihilation in the hadronic phase by measuring BFs, especially those binned by relative invariant momentum, $q_{\rm inv}$. Due to the large strength of the annihilation cross section at small $q_{\rm inv}$ as illustrated in Eq. (\ref{eq:sigmaA}), there will be a deficit of $p\bar{p}$ pairs at small relative momentum. The BF, which measures the relative number of opposite-sign vs. same-sign pairs, should then have a dip for $q_{\rm inv}\lesssim 100$ MeV/$c$. As this scale is lower than the thermal momentum, or other scales of the charge balance function, its strength can be readily separated from other physics, and thus unambiguously quantify the amount of annihilation in the hadronic phase.

To illustrate the efficacy of the strategy outlined above we compare calculations of BFs with and without annihilation for Pb+Pb collisions at $\sqrt{s_{NN}}=2.76$ TeV. Calculations are based on the methods from \cite{Pratt:2021xvg}: two-particle correlations are sourced and propagated assuming that local correlations are consistent with chemical equilibrium, according to charge susceptibilities from lattice calculations \cite{Borsanyi:2011sw}.  The balancing part of the correlations, whose strengths are fixed by charge conservation, are assumed to spread diffusively according to temperature-dependent diffusion constants, which are also determined by lattice calculations \cite{Aarts:2014nba,Amato:2013naa}. The model propagates these correlations using the hydrodynamic history of the collision, until $T_H$ is reached, at which point the correlations are projected onto hadronic degrees of freedom according to statistical arguments. Additional contributions from the evolution and decay of hadrons in the hadronic phase are then added to the correlation.

In the previous calculations cited above, annihilation was omitted. Here, annihilation is added, along with the inverse process. The resulting proton-antiproton balance functions, binned by relative azimuthal angle, relative rapidity and $q_{\rm inv}$, are defined by 
\begin{eqnarray}\label{eq:Bdef}
B(\Delta\phi)&=&\frac{1}{N_++N_-}\int dp_1dp_2\left\{
N_{+-}(p_1,p_2)-N_{-+}(p_1,p_2)\right.\\
\nonumber
&&\left.\hspace*{100pt}-N_{++}(p_1,p_2)-N_{--}(p_1,p_2)\right\}
\delta(\phi_1-\phi_2-\Delta\phi),\\
\nonumber
B(\Delta y)&=&\frac{1}{N_++N_-}\int dp_1dp_2\left\{
N_{+-}(p_1,p_2)-N_{-+}(p_1,p_2)\right.\\
\nonumber
&&\left.\hspace*{100pt}-N_{++}(p_1,p_2)-N_{--}(p_1,p_2)\right\}
\delta(y_1-y_2-\Delta y),\\
\nonumber
B(q_{\rm inv})&=&\frac{1}{N_++N_-}\int dp_1dp_2\left\{
N_{+-}(p_1,p_2)-N_{-+}(p_1,p_2)\right.\\
\nonumber
&&\left.\hspace*{100pt}-N_{++}(p_1,p_2)-N_{--}(p_1,p_2)\right\}
\delta(q_{\rm inv}(p_1,p_2)-q_{\rm inv}),\\
\nonumber
q^2_{\rm inv}(p_1,p_2)&=&\frac{1}{4}\left[\left(\frac{(p_1+p_2)\cdot(p_1-p_2)}{(p_1+p_2)^2}\right)(p_1+p_2) - (p_1-p_2) \right]^2.
\end{eqnarray}
Here, quantities $N_{++}(p_1,p_2), N_{--}(p_1,p_2), N_{+-}(p_1,p_2)$ and $N_{-+}(p_1,p_2)$ describe the number of pairs of the given charges with momentum $p_1$ and $p_2$. For example, $N_{+-}(p_1,p_2)$ represent the number of pairs with a positive particles having  momentum $p_1$ and a negative particle having momentum $p_2$. In this paper, the focus will be on BFs constructed using only protons and antiprotons. With this definition, $q_{\rm inv}$ is half the relative momentum in the pair's rest frame.

In the next section we review the model, with a focus on how baryon regeneration is incorporated into the hadronic simulation. The following section describes how annihilation and regeneration have been added to the model. Results and a summary comprise the subsequent sections.

\section{Theory and Model Overview}

Calculations for this study required several steps:
\begin{enumerate}
  \item The hydrodynamics code was run with initial conditions corresponding to the 0-5\% most central collisions of $\sqrt{s_{\rm NN}}=2.76$ TeV Pb+Pb collisions at the LHC. The temperature, flow, and stress-energy tensor were stored as a function the transverse spatial coordinate and proper time $\tau\equiv\sqrt{t^2-z^2}$. Boost invariance was assumed, implying that the evolution does not depend on spatial rapidity. This was the same evolution used in \cite{Pratt:2021xvg}. The hydrodynamic evolution was also analyzed to find the hyper-surface for transitioning into the hadron phase.

  \item Using the temperature evolution stored in (1), the value of the charge susceptibility matrix, $\chi_{ab}(x,y,\tau)$, and the diffusivity, $D_{ab}(x,y,\tau)$, were assigned for each space-time point according to lattice values \cite{Aarts:2014nba,Amato:2013naa} corresponding to the local temperature. The charge-charge correlation function was assumed to stay equilibrated, i.e. its strength was given by $\chi_{ab}$.  As described in \cite{Pratt:2017oyf,Pratt:2018ebf}, the fact that the overall charge-charge correlation integrates to zero, requires that the non-local correlation, which spreads according to the diffusive equation, must have a source term determined by the evolution of $\chi_{ab}$. Using the source function, the non-local charge-charge correlation function, $C_{ab}(x_a,y_a,x_b,y_b,\tau)$, was calculated as a diffusive equation. It was precisely this non-local part that becomes the balance function. The correlation functions  were represented by weighted pairs of charges in a Monte Carlo procedure. The pairs were assigned charges, e.g. $u,s$, and assigned weights which could be positive or negative based on the sign of the source term. Each charge $\delta q_a$ at some point passed through the hyper-surface boundary separating the hydrodynamic and microscopic descriptions. At that point, the charge stochastically created hadrons $\delta N_h$ \cite{Pratt:2017oyf,Pratt:2018ebf},
\begin{eqnarray}
\delta N_h&=n_h(T_c)\chi_{ab}^{-1}(T_H)Q_{ha}\delta q_b,
\end{eqnarray}
  where $\delta N_h$ is a hadron of type $h$, $n_h$ is the density of hadrons of that type at the hadronization temperature, and $\chi_{ab}(T_H)$ is the susceptibility. 

  Hadrons generated from two charges in a pair were tagged so that correlations between hadrons could be represented using hadrons from the same pair, thus avoiding combinatoric noise. The assignment included the weight of the pair. The hadrons from $\delta q_a$ and those from $\delta q_b$ were then propagated through the cascade, assuming that they  collided only with the background particles from the cascade, which are desribed in (3) below. The background particles did not themselves scatter, but this procedure effectively accounted for the small additional spatial spread of charge during the hadronic phase. The hadrons, $\delta N_{ha}$ and $\delta N_{hb}$ from the two charges, or their decaying descendants, were then paired with one another to calculate BFs, carrying over the original weights assigned to the pair $q_{ha},q_{hb}$. These correlations are referred to as type-I correlations, and were all generated from the source functions for the charge-charge correlation function during the hydrodynamic stage. This procedure explicitly ignores any correlation between the background particles and the hadrons coming from the sample charges. This approximation should be justified if the hadrons representing the correlation ultimately produce the same groups of particles, on average, that would be produced if they decayed 

  \item Using the hyper-surface information mentioned in (1), uncorrelated hadrons were emitted from the hyper-surface consistent with a thermalized system with viscous corrections. Resonances were populated using the same spectral functions employed by recent calculations of the SMASH model \cite{Petersen:2018jag,Staudenmaier:2017vtq}. These particles were then fed into a hadronic simulation, also referred to as a cascade. Hadrons collided according to resonant cross sections chosen to populate resonances consistently with the spectral functions. A 10 mb elastic $s-$wave cross section was also included. Unlike previous balance-function calculations performed by this group, baryon annihilation was added. By averaging 14400 events, each of which covered 2 units of rapidity with cyclic boundary conditions, the local stress-energy tensor and densities were found for individual species. From this information, kinetic temperatures, effective chemical potentials and collective velocities were found for specific radii, times and species.  These chemical potentials and temperatures in turn were then used to modify the annihilation cross section to effectively account for the inverse process. This involved repeating the simulations three times using fugacities calculated from previous runs, so that the fugacities converged. The theoretical basis and methods for taking into account the inverse process are described in more detail below. Because annihilation depends on these quantities, the calculation was repeated three times using chemical potentials and temperatures from the previous run. After three times the quantities had converged.

  \item A second contribution to BFs was calculated, using the final runs of (3) above. These correlations were those that were sourced during the cascade and are referred to as type-II. Type-II correlations include those from the decays of neutral hadrons into charged particles and those from baryon annihilation. This contribution was calculated in a brute-force manner by combining all final-state hadrons from (3) with one another. The type-II contributions were added to the type-I correlations to calculate BFs for all possible final species combinations. This involved adding the two BF numerators, then dividing by the final yields coming from the cascade. This procedure was tested in \cite{Pratt:2021xvg} by calculating the BFs integrated over all momentum (including outside the acceptance) and summing over all species. These tests verify that for any particles, the balancing electric charge matches the charge on the specific species. The procedure passed this test to better than 1 percent, consistent with the uncertainty expected from the Monte Carlo sampling. As was done in \cite{Pratt:2021xvg}, the acceptance of the ALICE detector was taken into account in a manner to be consistent with the ALICE BF analysis \cite{ALICE:2021hjb}. 
\end{enumerate}

This procedure was performed for three cases: first, without baryon annihilation; second, with annihilation but without baryon regeneration; and finally, with both annihilation and regeneration. Because annihilation only affects type-II correlations, the same type-I correlations were used for all three cases.

\section{Accounting for Baryon Regeneration During the Hadron Cascade}

When two nucleons annihilate into mesons, the mesons then predominantly decay into pions, with the typical number of outgoing pions being $\approx 5$. Thus, during the decays nucleon annihilations were all assumed to produce five mesons, with momenta chosen to reproduce the energy and momentum of the baryon pair and weighted by invariant phase space,
\begin{eqnarray}
dN&\sim& \left(\prod_{a}\frac{d^3p_a}{E_a}\right)\delta^4\left(P-\sum_a p_a\right).
\end{eqnarray}
The combination of meson species was chosen by randomly combining five quarks with five anti-quarks. Three of the five quarks, and three of the five antiquarks were taken from the constituent quarks of the annihilating baryons. The remaining quarks were chosen to be two up quarks 25\% of the time, two down quarks 25\% of the time and one of each for the remainder. The remaining antiquarks were chosen accordingly to conserve charge. The quarks were randomly combined with antiquarks to produce five pions. These choices are not particularly well motivated, but as long as charge is conserved, they will only negligibly affect the BF results shown here.

As pointed out by Shuryak and Rapp \cite{Rapp:2000gy}, the inverse process, five mesons combining into a baryon and antibaryon, should exactly cancel the annihilation process when a system is at equilibrium. Thus, if the hadron cascade is seeded by assuming a chemically-equilibrated system at the hyper-surface where the temperature is $T_c$, all annihilation along that hyper-surface should be canceled by baryon regeneration. Due to the rapid cooling of the system, however, local chemical equilibrium is lost at later times. If local kinetic equilibrium is maintained after the loss of chemical equilibrium, so that the momenta can be characterized by a collective velocity $\vec{u}$ and a local temperature $T$, then the density of each species $h$ can still be characterized by an ``effective'' chemical potential, $\mu_h$, defined by
\begin{eqnarray}
n_h&=&e^{\mu_h/T_H}n_{h,{\rm eq.}}(T),
\end{eqnarray}
where $n_{h,{\rm eq.}}(T)$ is the density of a species at equilibrium.

In a completely equilibrated system, the chemical potential for a particle and anti-particle would be equal and opposite, and would approach zero in a system with little or no net charge, such as the matter produced at mid-rapidity in a high-energy collision at the LHC. Here, a system with no net charge would then lead to the effective potentials being equal for a particle and its anti-particle. Such effective potentials have already been extracted from heavy-ion collisions going back to SPS collisions \cite{Greiner:1993jn,Pratt:1998gt,Pan:2014caa}, where typical effective chemical potentials for pions were in the range of 70 MeV and near 350 MeV for protons at final breakup, when kinetic temperatures are in the neighborhood of 100 MeV.

These effective chemical potentials allow one to express the inverse rate for baryon regeneration in terms of the annihilation rate by introducing an appropriate combination of suppression factors, according to the following expressions:
\begin{eqnarray}\label{eq:inverserate}
R(5~{\rm mesons}\rightarrow B,\bar{B})&=&R(B,\bar{B}\rightarrow 5~{\rm mesons})
S_\mu(\vec{\mu},\vec{T})S_T(\vec{\mu},\vec{T}),\\
\nonumber
S_\mu(\vec{\mu},\vec{T})&=&\exp\left\{
-\frac{\mu_B}{T_B}-\frac{\mu_{\bar{B}}}{T_{\bar{B}}}+\sum_m\frac{\mu_m}{T_m}
\right\},\\
\nonumber
S_T(\vec{\mu},\vec{T})&=&\exp \left\{
E/2T_B+E/2T_{\bar{B}}-\sum_m(E/5)T_m
\right\}.
\end{eqnarray}
where $\mu_{B},\mu_{\bar{B}},T_B$ and $T_{\bar{B}}$ depend on the specific baryons being annihilated, and the sum over $m$ represents the sum over the 5 mesonic products. Here, $E$ is the energy being converted from one sector to the other, i.e. the summed energies of the baryon and anti-baryon in the fluid frame. To simplify the expression, it was assumed that the energy in the mesonic sector was evenly split among the sectors for each species.

When the suppression factors $S_\mu$ and $S_T$ equal unity, recombination exactly cancels the annihilation. When particles in the baryonic sector have a higher net chemical potential of the five mesons, or if the baryons have lower temperature, the recombination only partially cancels the annihilation.

To understand Eq. (\ref{eq:inverserate}) we consider two systems $a$ and $b$, with different chemical potentials and temperatures. The ratio of rates to inverse rates should equal the ratio of the number of available states, $e^{\Delta S}$, which is taken to be the ratio of the Boltzmann factors before and after the annihilation.  $\Delta S$ is the entropy gained or lost by the reaction, and is then given by
\begin{equation}
\Delta S=E/T_a-E/T_b-\sum_{h_a}\mu_a/T_a+\sum_{h_b}\mu_b/T_b.
\end{equation}
Because more massive particles tend to cool more quickly \cite{Pratt:1998gt}, the temperature difference also increases the regeneration rate. The differential cooling can be understood by considering a collisionless Hubble expansion. In that limit the kinetic temperature of non-relativistic particles falls as $1/\tau^2$, whereas the kinetic temperature of massless particles fall as $1/\tau$. Further, as the system cools, pions tend to filter through the more massive particles, and their outward collective velocity separates from that of the heavier particles. 

Because one can estimate the regeneration rate using the suppression factors $S_\mu$ and $S_T$, we can circumvent implementing the $5\rightarrow 2$ processes in simulation by simply reducing the annihilation rate by the factor $[1-S_\mu(\vec{\mu},\vec{T})S_T(\vec{\mu},\vec{T})]$.  One therefore has
\begin{eqnarray}\label{eq:invrate}
R(B,\bar{B}\rightarrow 5~{\rm mesons})&\rightarrow&
R_{\rm vacuum}(B,\bar{B}\rightarrow 5~{\rm mesons})[1-S_\mu(\vec{\mu},\vec{T})S_T(\vec{\mu},\vec{T})].
\end{eqnarray}
This strategy was applied in \cite{Pan:2014caa} for a schematic model (not a cascade). The $5\rightarrow 2$ inverse process has been implemented in a cascade, but only for proton-antiproton annihilations \cite{Garcia-Montero:2021haa}, ignoring annihilations of other baryons.

One challenge in implementing this approach is finding temperatures and effective chemical potentials for all species at all points in space time. For mesons, one need only find these quantities for pions and kaons, but there is a large array of baryons that might annihilate, some of which are too rare to gain sufficient statistics to extract thermodynamic quantities. Thus, the quantities were extracted for the lowest lying baryon flavor octet and decuplet, whereas higher-mass baryons were assigned the same $T$ and effective chemial potential as the lower lying states of the same strangeness and total isospin. 

As noted previously, calculations were based on a cascade where particles were generated from a hydrodynamic model of central (0-5\% centrality) $\sqrt{s}_{nn}=2.76$ TeV Pb+Pb collisions, and the same approach was used to compute charge balance functions as was employed in Ref.~\cite{Pratt:2021xvg}. The interface temperature was chosen to be $T_c=155$ MeV.  Because of the assumed boost invariance \cite{Bjorken:1982qr}, the temperatures and chemical potentials only needed to be determined as functions of the transverse coordinates, $x$ and $y$, and the proper time, $\tau$.

At each time step, the stress energy tensor and densities were calculated for particles within the cells by averaging $10^4$ collisions together. The stress-energy tensor was calculated for each species $h$ by considering the particles within a given cell with volume $V$,
\begin{eqnarray}{}
T_h^{\mu\nu}(x,y)&=&\frac{1}{V}\sum_{p_h\in V}\frac{p_h^\mu p_h^\nu}{E_h}.
\end{eqnarray}
This was sufficient to determine the velocity of the fluid $U^\mu$, i.e. the velocity of the frame where $T_h^{0i}=0$ for $i=1,2,3$, and the energy density in the rest frame of the fluid for the species $h$, $\epsilon_h$.
The local density of the rest frame, $n_h$, was then found with $U_0n_h=N_h/V$. Once $\epsilon_h$ and $n_h$ were known, one numerically solved for the temperature, $T_h$, and for the effective chemical potenial, $\mu_h$, in each cell. These values were averaged over radial slices to find the temperatures and chemical potentials as a function of the radius, $r=\sqrt{x^2+y^2}$, for specific proper times $\tau$. Because the densities depended on annihilation, and because annihilation depended on the chemical potentials extracted from the densities, calculation of the thermodynamic quantities was repeated three times. In each calculation, thermodynamic quantities from the previous run were used to calculate the regeneration rates. After repeating the calculation three times, results had converged.

\section{Results}

Before exploring the consequences of this analysis for the interpretation of charge balance functions, we first discuss temperatures, densities, collective radial velocities and effective chemical potentials obtained using the procedure described above.

\begin{figure}
\centerline{
\includegraphics[width=0.333\textwidth]{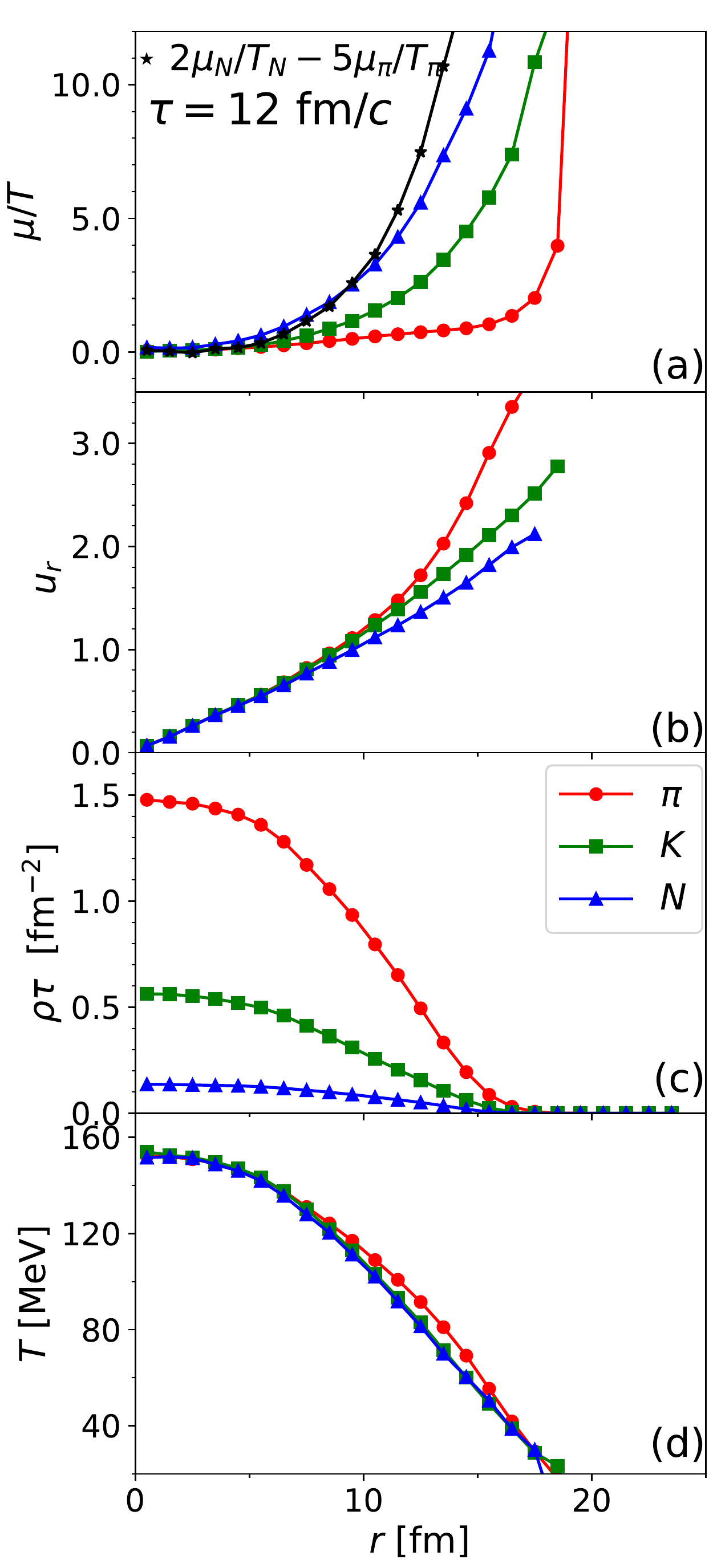}
\includegraphics[width=0.333\textwidth]{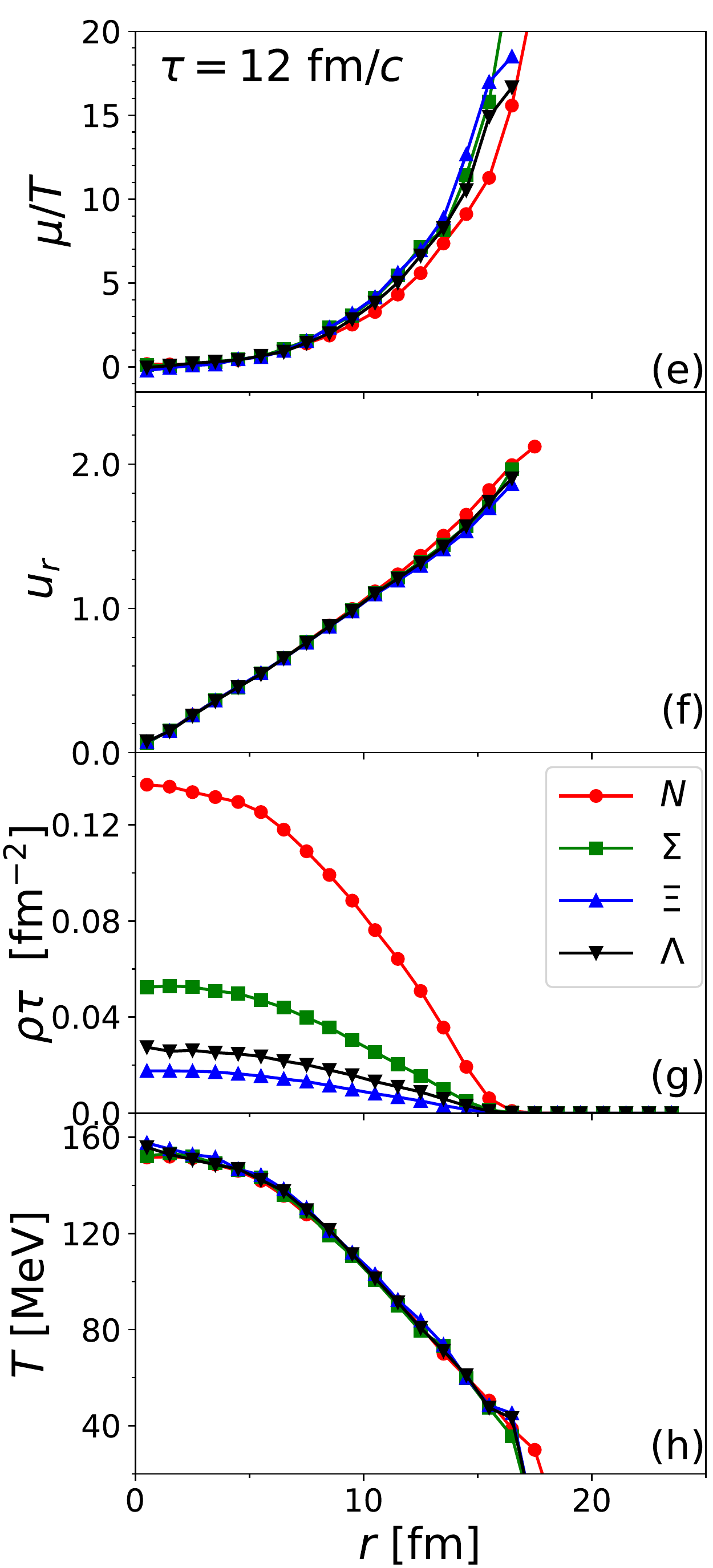}
\includegraphics[width=0.333\textwidth]{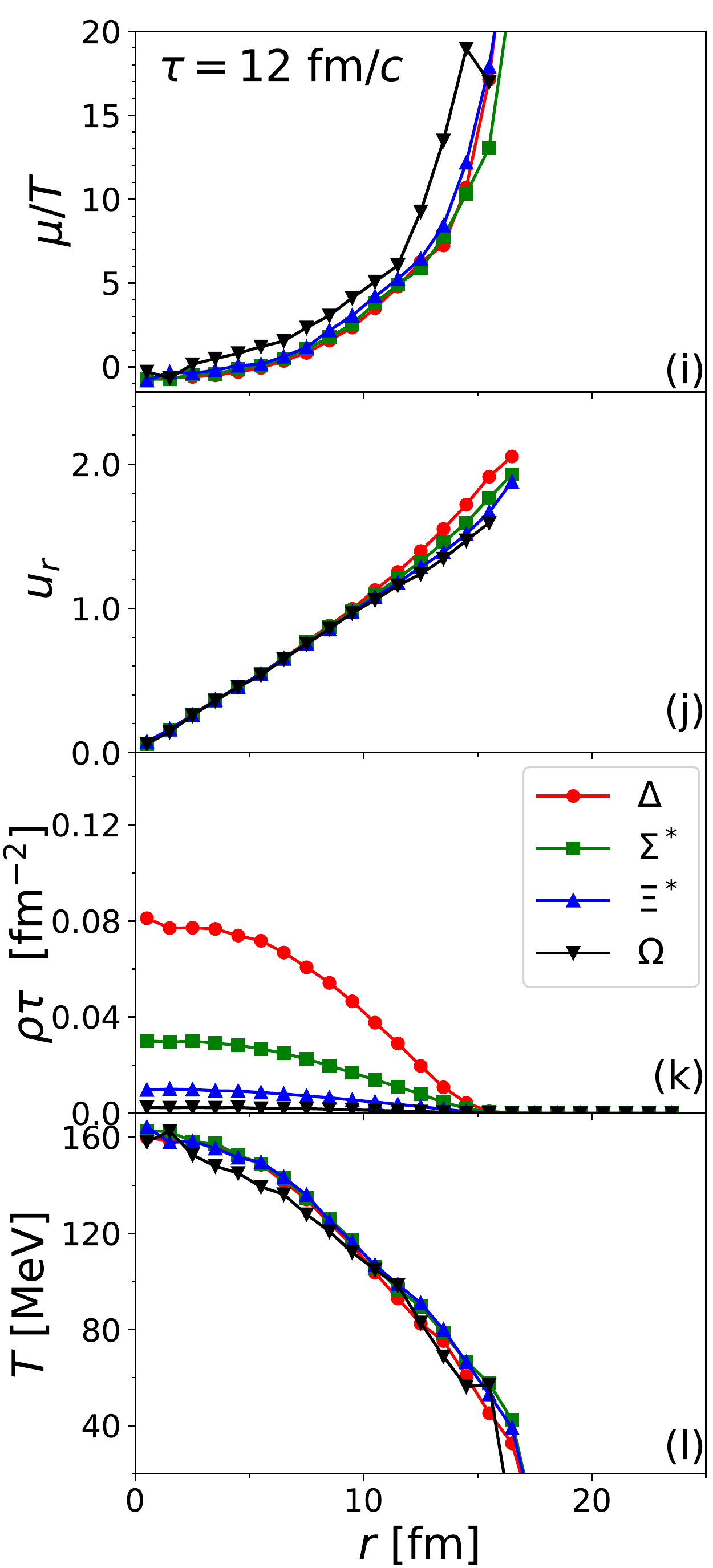}
}
\caption{\label{fig:mutvsr_tau12}
The left-side panels (a-d) show the effective thermodynamic quantities for pions, kaons and nucleons as a function of radius for the time $\tau=12$ fm/$c$, which is immediately after the hydrodynamic description has been completely replaced by the hadronic cascade. The local temperature $T$, number density $\rho$, collective radial velocity $u_r$ and chemical potentials are presented. The same quantities are displayed for spin-1/2 and spin-3/2 baryons in the central (e-h) and right-side (i-l) panels. At this time the central region has left the hydrodynamic stage and quantities are still nearly thermalized. For nucleon-nucleon annihilation, $2\mu_N/T_N-5\mu_\pi/T_\pi$ is the relevant combination of chemical potentials for determining whether recombination is important. When this combination is small, most annihilation processes are cancelled.}
\end{figure}

\begin{figure}
\centerline{
\includegraphics[width=0.333\textwidth]{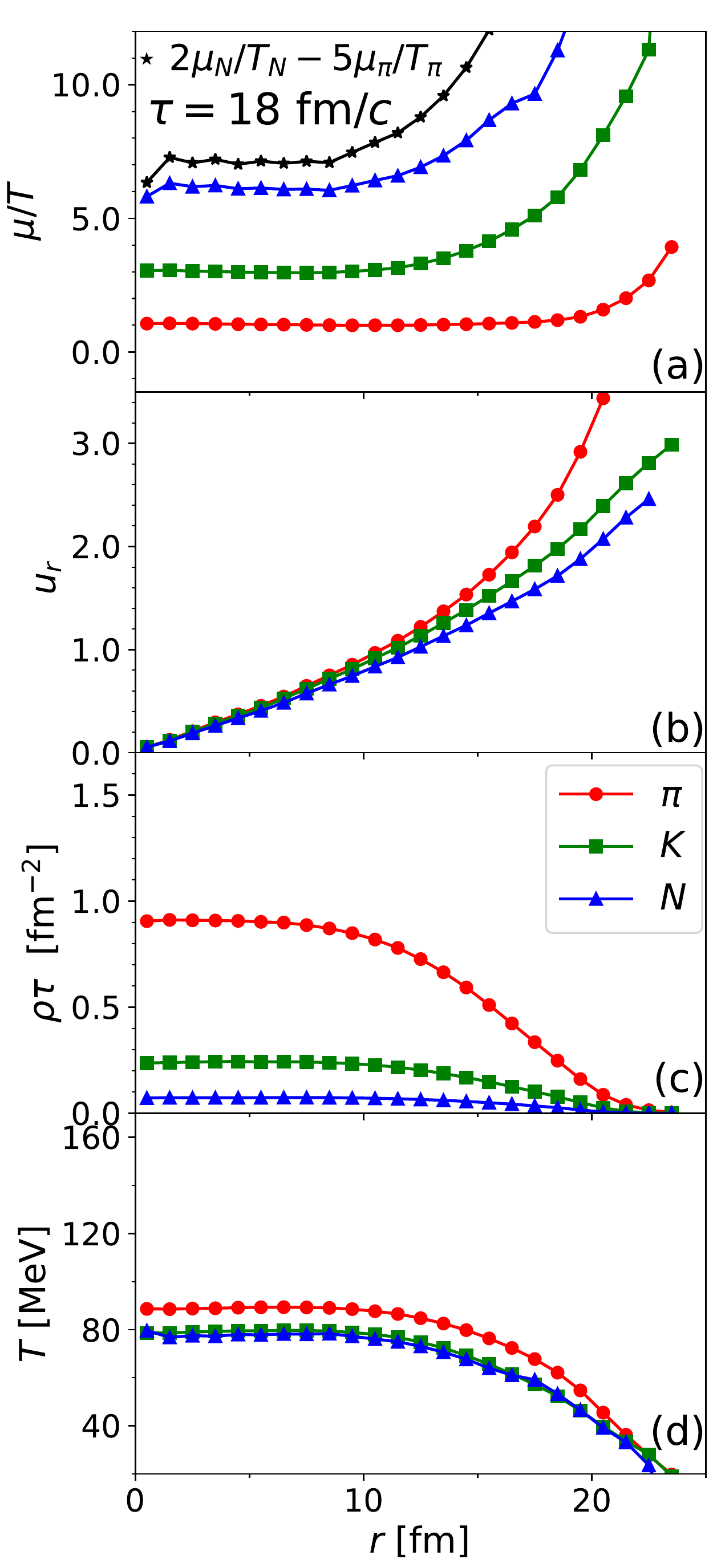}
\includegraphics[width=0.333\textwidth]{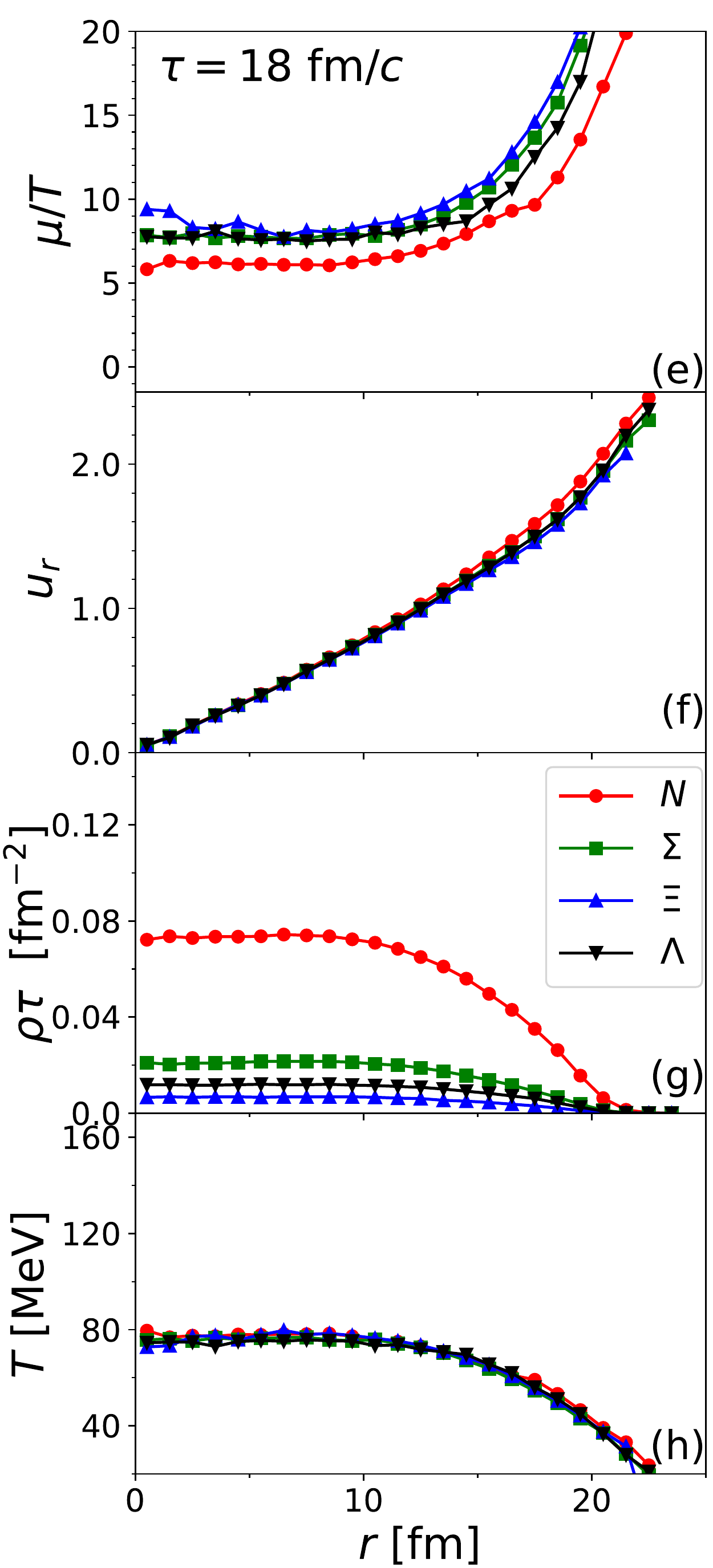}
\includegraphics[width=0.333\textwidth]{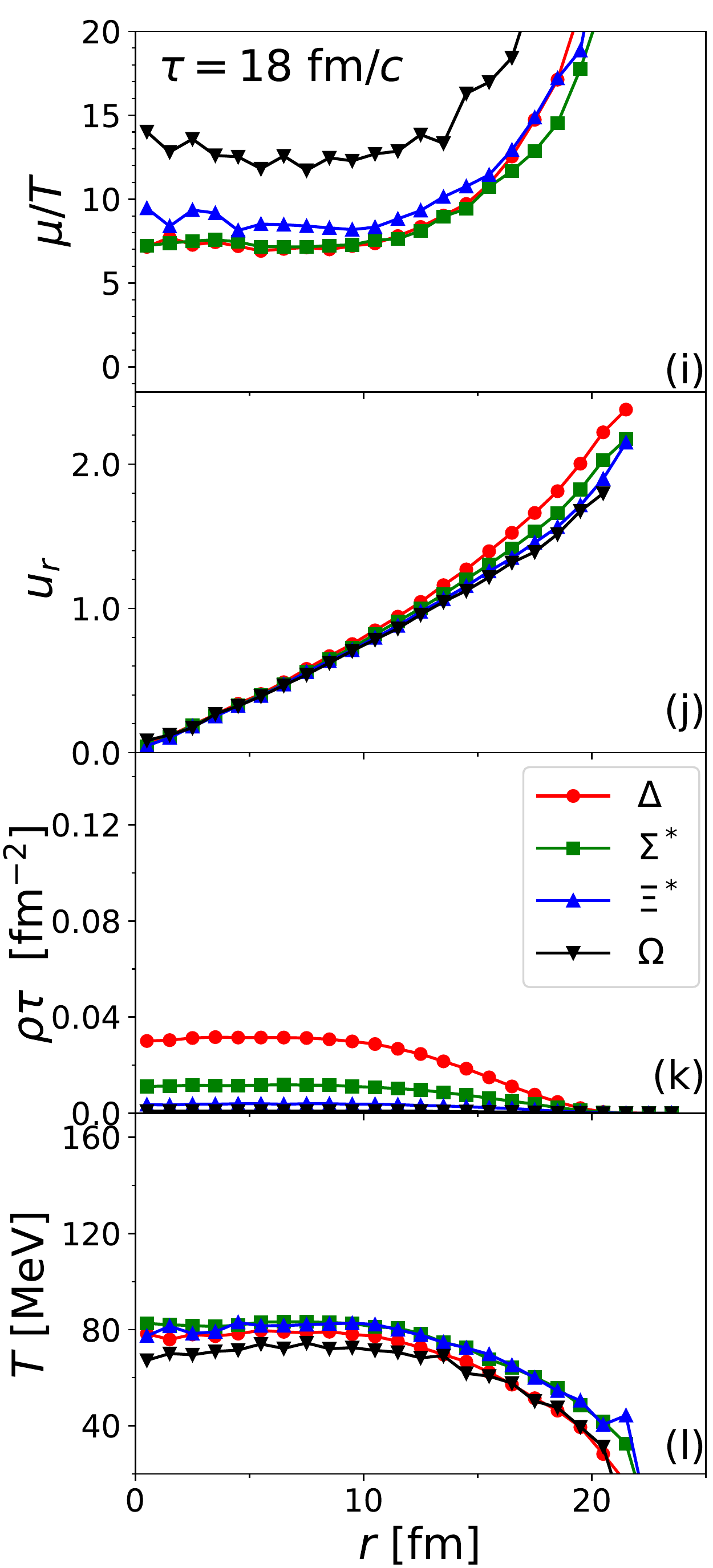}
}
\caption{\label{fig:mutvsr_tau18}
Thermodynamic quantities are shown at a time $\tau=18$ fm/$c$, a time when many of the emitted hadrons are experiencing their final interactions. By that time the effective chemical potentials have grown substantially, and the combination, $2\mu_N-5\mu_\pi$ is large, which in turn means that the baryon recombination rates are negligible by that time. The right-side panels (i-l) show the thermodynamic quantities for other baryon species. They resemble those for nucleons.}
\end{figure}

These quantities are displayed in Figs.~\ref{fig:mutvsr_tau12} and \ref{fig:mutvsr_tau18} at both $\tau=12$ fm/$c$ and $\tau=18$ fm/$c$. The calculation used $10^4$ events covering 10 units of spatial rapidity. The hyper-surface representing $T=T_c=155$ MeV/$c$ ceased existing just before $\tau=12$ fm/$c$, meaning that by this time, most of the matter has barely cooled below $T_c$ and the effective chemical potentials are close to zero.  By $\tau=18$ fm/$c$ roughly half of the final-state particles have been emitted, and at that time the chemical potentials are large.

Figures \ref{fig:mutvsr_tau12} and \ref{fig:mutvsr_tau18} also show the degree to which various species share the same collective velocities and local temperatures.  As thermal equilibrium is lost, local kinetic temperatures for heavier particles in Fig.~\ref{fig:mutvsr_tau18} fall below those of pions and the collective flow velocity begins to vary by species as lighter particles begin to flow through the heavier particles (an effect sometimes referred to as the ``pion wind" \cite{Pratt:1998gt}). This calls into question the use of Eq. (\ref{eq:inverserate}), which presumes species share the same collective velocity.  Fortunately, however, the differences in the temperatures, chemical potentials and flows in Figs.~\ref{fig:mutvsr_tau12} and \ref{fig:mutvsr_tau18} are rather modest, and in those regions where the difference is larger the recombination rates would so small in any case that any temperature or flow difference is inconsequential. 

\begin{figure}
\centerline{
\includegraphics[width=0.9\textwidth]{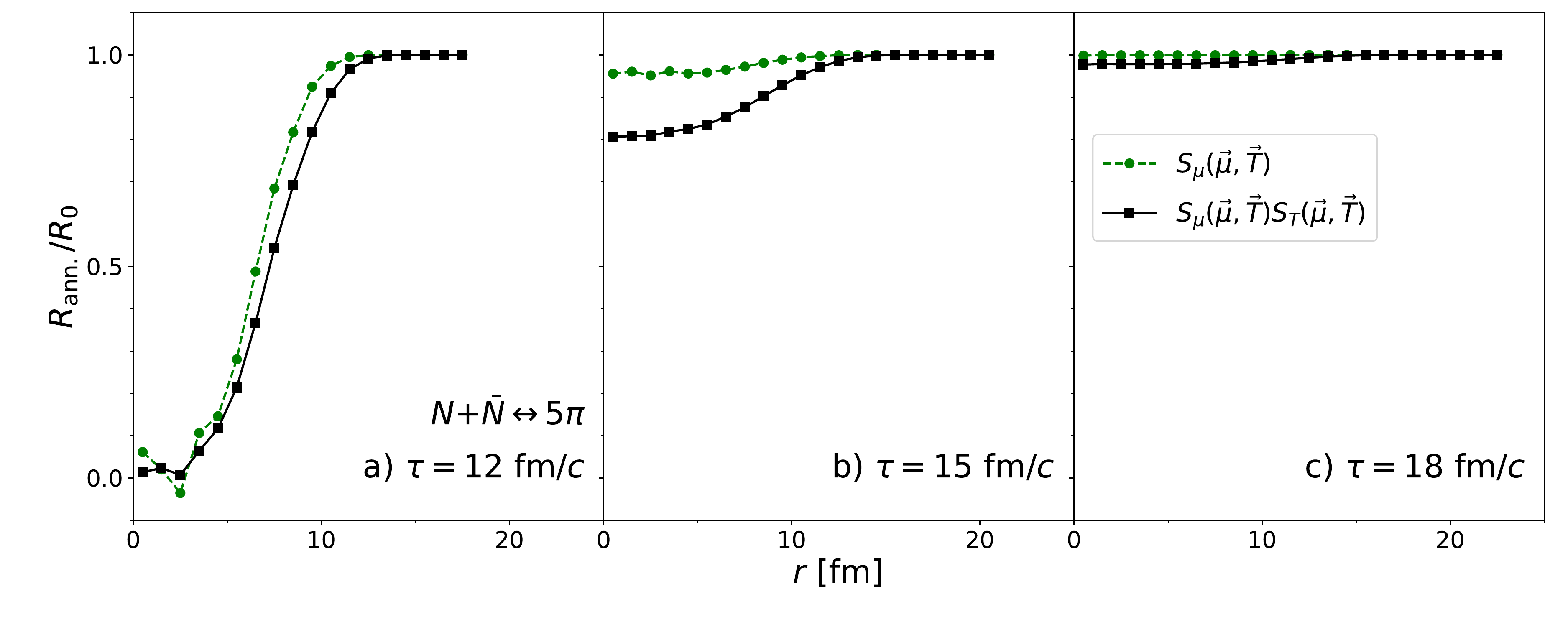}
}
\caption{\label{fig:KillRatio}
Annihilations are suppressed by the factors illustrated here. Immediately after the transition from hydrodynamics, nearly all annihilations are suppressed, whereas toward the end of the reaction nearly all annihilations proceed. Suppression due to the difference in chemical potentials, $(1-S_\mu(\vec{\mu},\vec{T}))$, and the full suppression factor, $(1-S_\mu(\vec{\mu},\vec{T})S_T(\vec{\mu},\vec{T}))$ are both shown.}
\end{figure}

Annihilation suppression factors are shown in Fig. \ref{fig:KillRatio}.  The factors are near zero shortly after the matter cools below $T_c$, but approach unity by the time matter is emitted at $\tau\approx 18$ fm/$c$. For the central collisions modeled here, approximately 18\% of baryons were annihilated if the regeneration was neglected, i.e. if $S_\mu(\vec{T},\vec{\mu})$ and $S_T(\vec{T},\vec{\mu})$ were set to zero. After incorporating the suppression factors in Fig. \ref{fig:KillRatio} roughly 12\% of baryons were annihilated during the cascade.

Annihilation during the cascade only affects the type-II contributions to the charge balance function. The type-II contributions were calculated for three cases: without annihilation, with annihilation but without suppression, and finally, with annihilation including the suppression factors. When baryon annihilation is not invoked, the type-II contributions to the proton-antiproton BF is negligible. For each case, 9600 cascade events were analyzed. In order to cover the spread of correlation, cyclic boundary conditions were employed at spatial rapidities, $\eta_{\rm min,max}=\mp 5$. Give that experimental coverage for identified particles are in the range of $|\eta|<0.8$, and given that efficiencies for the calculation were perfect, the statistical noise in these calculations is similar to what an experiment might measure with $\approx 10^5$ events. High statistics runs with the STAR detector at RHIC or at ALICE at the LHC (Large Hadron Collider) might ultimately have more than ten times these statistics. 

\begin{figure}
\centerline{
\includegraphics[width=0.333\textwidth]{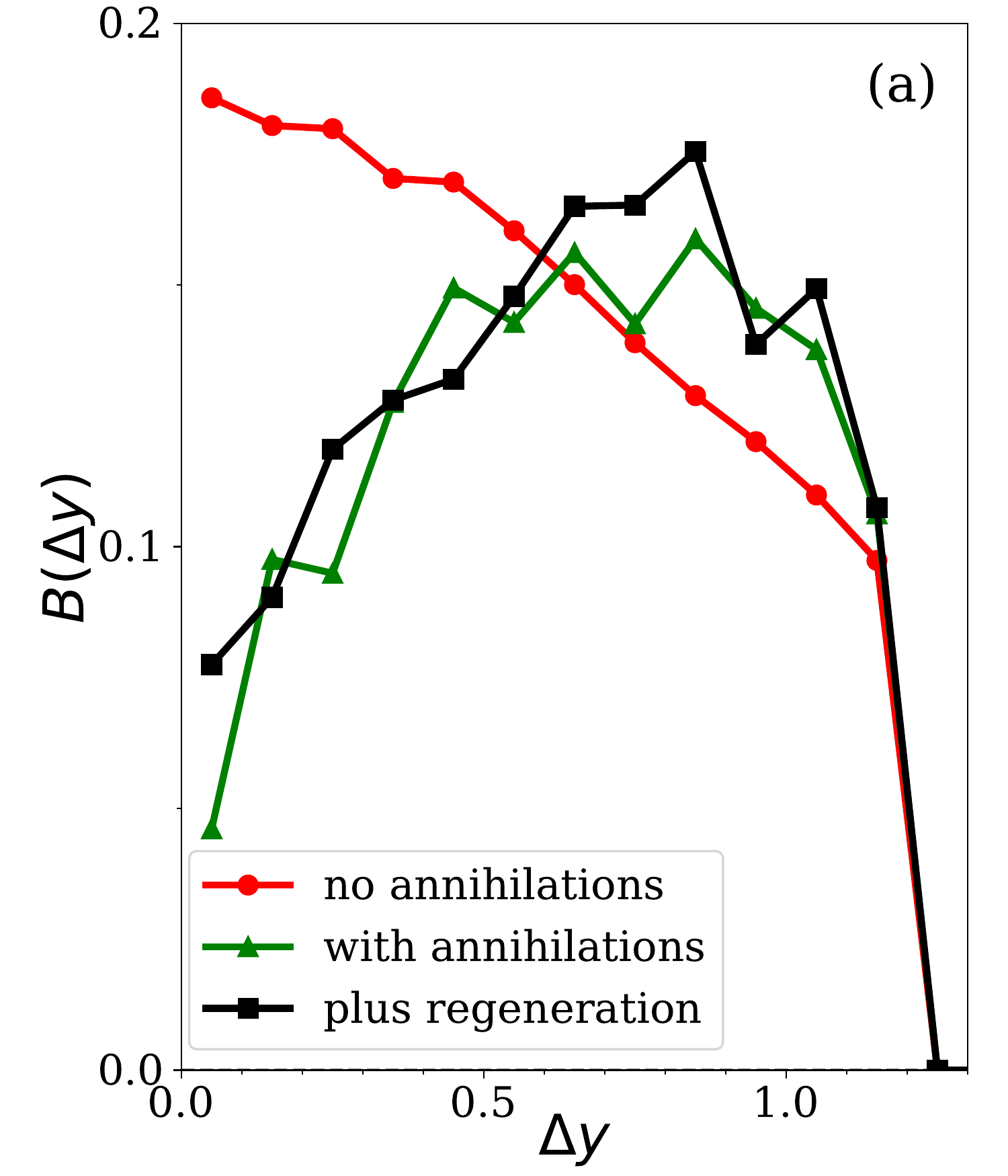}
\includegraphics[width=0.333\textwidth]{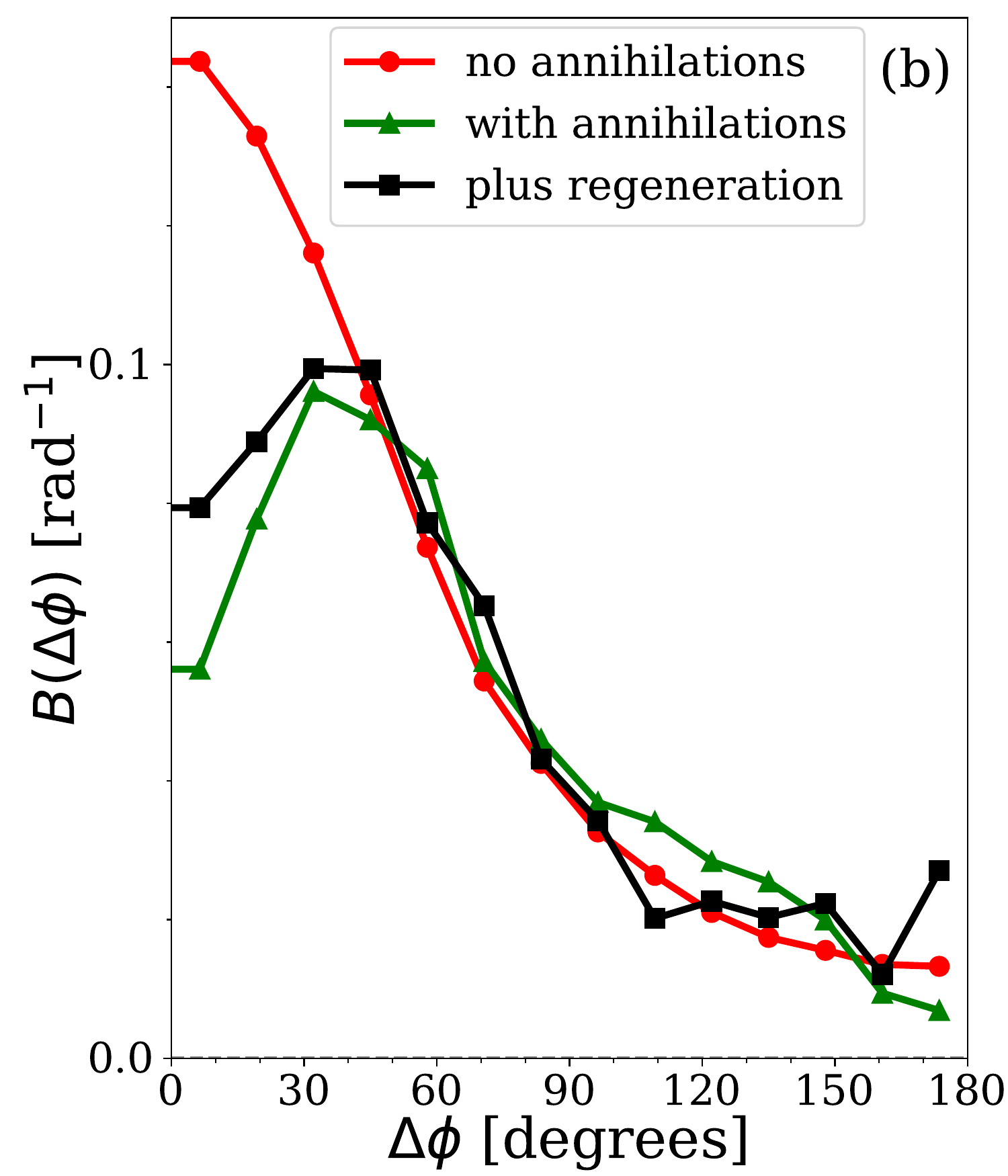}
\includegraphics[width=0.333\textwidth]{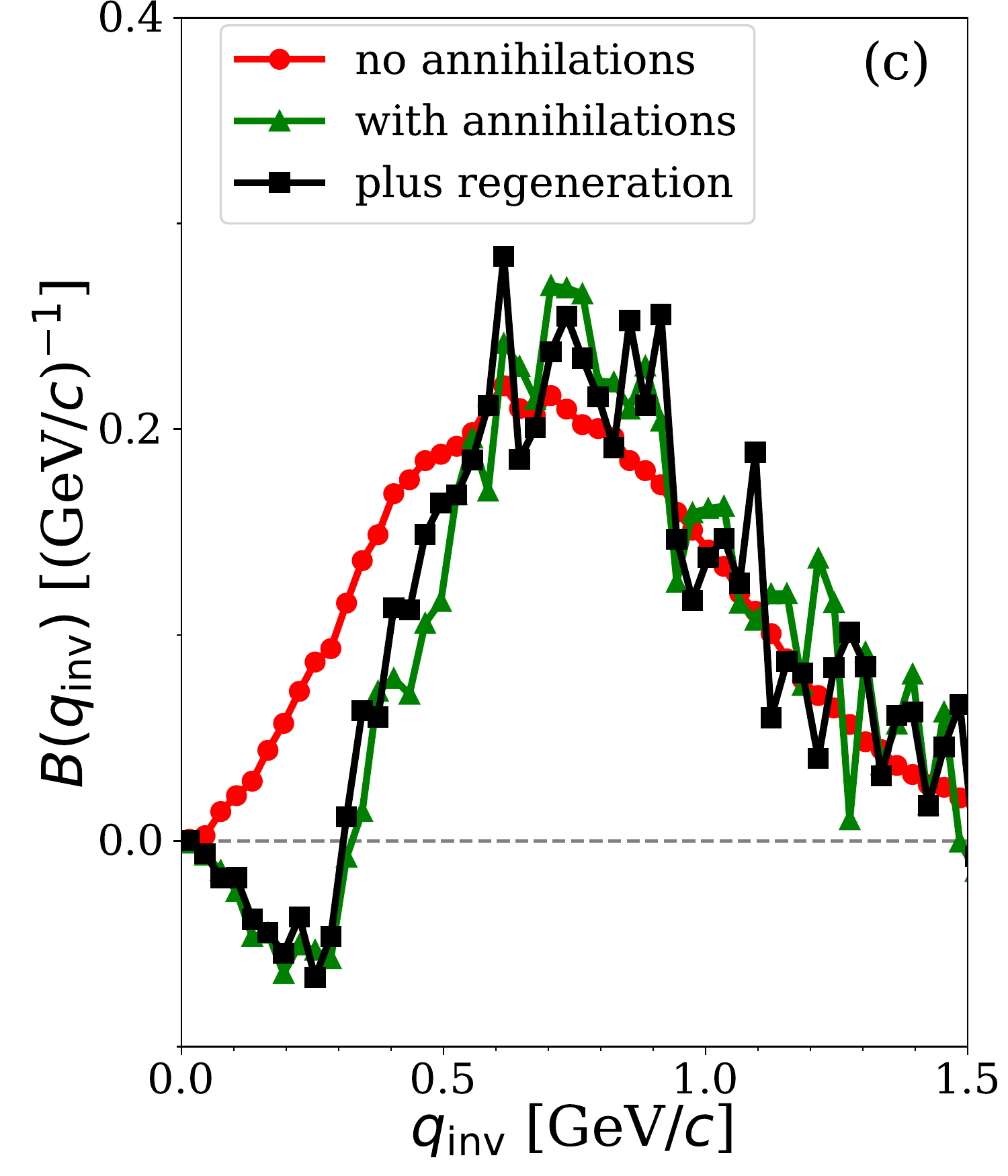}
}
\caption{\label{fig:BFs}Charge balance functions using protons and antiprotons binned by (a) relative rapidity, (b) relative azimuthal angle, and (c) relative invariant momentum, $q_{\rm inv}$. An acceptance cutoff constrains the balance fucntion to $\Delta y> 1.2$. Binning in $q_{\rm inv}$ is especially useful for identifying baryon annihilation in the hadronic stage.
}
\end{figure}
BFs using protons and antiprotons for the three cases are shown in Fig. (\ref{fig:BFs}). As expected, annihilation provides a significant dip at small relative rapidity, $\Delta y$, relative azimuthal angle, $\Delta\phi$ or the invariant relative momentum,
\begin{eqnarray}
q^2_{\rm inv}&=&
-(p_1-p_2)^2/4+[(p_1-p_2)\cdot P]^2/4P^2,
\end{eqnarray}
where $P^\mu=p_1^\mu+p_2^\mu$. With this definition $q_{\rm inv}$ is half the relative momentum in the rest frame of the pair. Of these three choices, $q_{\rm inv}$ provides the clearest means to view the contribution. This was expected given the large annihilation cross section for low relative momentum, and the fact that collective flow focuses the effects of annihilation toward smaller relative momentum. For charge balance correlations that depend only on relative position, thermal motion would lead to BFs that scale as $q_{\rm inv}^2$ for small relative momentum due to phase space. Aside from annihilation, there is no reason to expect the negative dip in Fig. \ref{fig:BFs}. Because the annihilation contribution is concentrated at small $q_{\rm inv}$ it is rather easy to identify it. The width of the broader type-I contribution might easily change, as it depends on details of how charge is created and diffuses during the hydrodynamic stage, but it should be rather featureless, and by fitting the entire form, it should be relatively straight-forward to quantitatively identify the contribution from annihilation, and therefore to quantitatively constrain the amount of annihilation in the hadronic stage. 

In addition to better understanding the amount of baryon annihilation in the latter stage of the collision, it is imperative to understand how proton-antiproton BFs, particularly those binned by $\Delta y$ and $\Delta\phi$, are distorted by annihilation. The width of the proton-antiproton BF in $\Delta y$ provides critical insight into whether quarks are produced early, within the first fm/$c$, of a heavy-ion collisions. This observable provides the field's best hope for understanding whether a chemically equilibrated QGP was indeed realized early in the reaction's evolution. By constraining the annihilation contribution by analyzing $B(q_{\rm inv})$, one can more confidently interpret the broader structure of the proton-antiproton BF.

\section{Summary and Conclusions}

There are reasons to doubt the accuracy of the model calculations presented here. The large annihilation cross section might change in the environment where several other particles might exist within the characteristic distance of $\sigma_{\rm annihilation}$. Further, the phenomenological annihilation cross section was motivated by $p\bar{p}$ annihilation data, and might significantly differ for other species. One or both of the annihilating species would likely be a neutron, a resonance like the $\Delta$, or a hyperon. But, even though the model calculations presented here have significant uncertainty, the conclusion that the annihilation contribution can be constrained by observing the baryon-antibaryon BFs binned by $q_{\rm inv}$ is robust. For very high-energy collisions, where particles and particles are produced with nearly equal probability, the main physical source of competing correlation comes from final-state interactions. This was studied in detail in \cite{Pratt:2022xbk}, where it was found that final-state interactions distort BFs only for $q_{\rm inv}<50$ MeV/$c$, so this competing contribution is easily separable. One might even better understand other annihilation cross sections through measurement of BFs involving other baryons, such as $\Lambda$s. 

If the calculations presented here for proton-antiproton BFs binned by $q_{\rm inv}$ indeed match measurements from the LHC, it would confirm that net baryon annihilation in the hadronic phase is near 12\% for central Pb+Pb collisions at the highest energies. Measurements of the ALICE Collaboration show the proton to pion ratio in Pb+Pb collisions falling by approximately 15\% as centralities change from mid-central to most central \cite{ALICE:2013wgn}. If only 12\% of baryons annihilate in the most central collisions, and given that there must also be some annihilation in mid-central collisions, it would seem that baryon annihilation is unlikely to explain more than half of this trend.

BFs binned by relative rapidity provide insight into the chemical evolution of the super-hadronic matter produced in heavy-ion collisions. The calculations displayed in Fig. \ref{fig:BFs} were based on local chemical equilibrium being maintained from very early times. This allowed diffusion to produce wide $p\bar{p}$ BFs. Here, ``wide'' is relative to the thermal spread expected if two balancing charges are emitted close to one another. The characteristic thermal spread is $\approx 0.4$ units of rapidity. However, one can see that the BFs binned by relative rapidity in Fig. \ref{fig:BFs} are significantly broader than the thermal spread. If instead of the early production used here, quark production had taken place later, as one might have expected for a long-lived gluon plasma, balancing charges would have only spread by a few tenths of a unit of spatial rapidity and the BF widths would be of the order 0.5 units of rapidity. Thus, the observation of $p\bar{p}$ with spreads in rapidity $\gtrsim 1$ unit suggest early production of charge \cite{Pratt:2015jsa}. However, because one could adjust the narrowness with annihilation, it is imperative to accurately account for annihilation. Fortunately, as shown here, the characteristic scale of the dip in the BF from baryon annihilation is smaller than the thermal scale, which makes it readily identifiable, especially when binned by $q_{\rm inv}$. Once the shape of the proton-antiproton BF binned by $q_{\rm inv}$ is understood, then one can confidently state the contribution from annihilation to the BF binned by relative rapidity. The dip in the BFs seen in Fig. \ref{fig:BFs} due to annihilation is not unlike what has been seen in measurements from the ALICE collaboration, which are not yet published except as part of a thesis \cite{Pan:2019kqo}. Finalized results, and results from analyses with much higher statistics may be available within the next few years, both from the ALICE Collaboration at the LHC and from the STAR Collaboration at RHIC.

Thus, for both motivations listed above, it would be crucial to provide high-statistics BFs constructed from protons and antiprotons. The BFs should be binned by all three measures of relative momentum, $\Delta y$, $\Delta\phi$ and $q_{\rm inv}$. Once these measurements are published, the question of baryon annihilation in the hadronic stage should be largely settled. Further, once the contributions from annihilation to proton-antiproton BFs is clarified, one can embark on a detailed comparison of BFs from experiment to BFs from the model presented here. This would then enable quantitatively addressing the question of whether chemical equilibrium was established early ($\tau<1.0$ fm/$c$) in the light quark sector.

\begin{acknowledgments}
S.P., D.O. and C.P. were supported by the Department of Energy Office of Science through grants no. DE-FG02-03ER41259, no. DE-FG02-00ER4113 and no. DE-SC0020633 respectively.
\end{acknowledgments}

\end{document}